\documentclass[aps,showpacs,nofootinbib]{revtex4}

\usepackage[utf8]{inputenc}
\usepackage{color}
\usepackage{amsmath,amssymb}
\usepackage{amsfonts}
\usepackage{verbatim}
\usepackage{makeidx}
\usepackage{hyperref}
\usepackage{slashed}
\usepackage[vcentermath]{youngtab}
\usepackage{float}
\usepackage{booktabs}
\usepackage{graphicx}
\usepackage{mathrsfs}
\usepackage{bm}
\usepackage{braket}
\usepackage{url}
\usepackage{etoolbox}
\apptocmd{\sloppy}{\hbadness 10000\relax}{}{}
\allowdisplaybreaks[3]
 \newcommand{\be}{\begin{equation}}
   \newcommand{\ee}{\end{equation}}
     \newcommand{\bea}{\begin{eqnarray}}
   \newcommand{\eea}{\end{eqnarray}}

\usepackage[normalem]{ulem}  
\usepackage{color}
\renewcommand\sout{\bgroup \color{red} \ULdepth=-.5ex \ULset}

\begin{document}

\title{Radiative and meson decays of $Y(4230)$ in flavor SU(3) }
\author{Luciano Maiani}
\author{Antonio D. Polosa}
\author{Ver\'onica Riquer}
\affiliation{Dipartimento di Fisica and INFN,  Sapienza  Universit\`a di Roma\\ Piazzale Aldo Moro 2, I-00185 Roma, Italy}

\date{\today}

\begin{abstract}
The charmonium-like exotic states $Y(4230)$ and the less known $Y(4320)$, produced in $e^+e^-$ collisions, are sources of positive parity exotic 
hadrons in association with photons or pseudoscalar mesons. We analyze the radiative and pion decay channels in the compact tetraquark scheme, with 
a method that proves to work equally well in the most studied  $D^*\to \gamma/\pi+D$ decays. The decay of the vector $Y$ into a pion and a $Z_c$ state 
requires a flip of charge conjugation and isospin that is  described appropriately in the formalism used. Rates however are found to depend on the fifth power of pion momentum
which would make the final states $\pi Z_c(4020)$ strongly suppressed with respect to $\pi Z_c(3900)$. The agreement with BES III data would  be
improved considering the  $\pi Z_c(4020)$ events to be fed by the tail of  the $Y(4320)$ resonance under the $Y(4230)$. These results should renovate the interest in
further clarifying the emerging experimental picture in this mass region. 
\end{abstract}


  \maketitle
\let\endtitlepage\relax

\section{Introduction}\label{introd}

The study of final states in $e^+ e^-$ high energy annihilation, with the pioneering contributions by BaBar, Belle and BES collaborations, has opened the way to the new spectroscopy of exotic hadrons.

The so-called $Y$ states, unexpected charmonium-like states created by the initial lepton pair, are efficient sources of positive parity exotic hadrons produced in association with one photon, pion or K meson. 

Decays of the lightest $Y$ states, such as  $Y(4230)$ into $\gamma/\pi/K + X/Z$,  extensively studied by the BES III Collaboration, have provided precious informations on properties and quantum numbers of the lightest, $J^P=1^+$ exotic states (see e.g.~\cite{changzheng}), the latest result being the observation of the first, hidden charm, open strangeness $Z_{cs}(3985)$, produced in association with a charged K meson in~\cite{Ablikim:2020hsk} (exotic hadrons are extensively reviewed in~\cite{Ali:2019roi,Chen:2016qju,Esposito:2016noz,Ali:2017jda,Guo:2017jvc,Lebed:2016hpi,Olsen:2017bmm}).

In this note, we adopt the compact tetraquark model for $X(3872),~ Z_c(3900),~Z_c(4020)$, as $S$-wave tetraquarks~\cite{Maiani:2004vq,Maiani:2014aja,Maiani:2016wlq}, and for $Y$ states, as $P$-wave tetraquarks~\cite{ Maiani:2014aja,Ali:2017wsf},
to  study radiative and pionic decays of $Y(4230)$ 
\bea
&&Y(4230) \to \gamma +X(3872)\label{gamdec}\\
&&Y(4230) \to \pi+ Z_c(3900)/Z_c(4020),\label{pidec}
\eea 
observed  by BES III in the reactions
\be
e^+ e^- \to Y(4230) \to \pi+ Z_c(3900)/Z_c(4020)~{\rm or}~ \gamma +X(3872)\label{resprod}
\ee
 
 For a $Y$ resonance of valence composition $[cu][\bar c\bar u]$ or $[cd][\bar c\bar d]$, the photon in~\eqref{gamdec} is emitted from the light quark or antiquark. Decay \eqref{pidec} arises from the elementary transitions
\be
u\to  d ~\pi^+\quad {\rm or}\quad \bar d \to \bar u~\pi^+
\ee 
and similar for $\pi^-$ and $\pi^0$. The same transitions are operative in $D^*\to \gamma/\pi +D$ decays~\cite{Casalbuoni:1996pg}.
   
Our results for $Z_c(3900)$ and $X(3872)$ are in quantitative agreement with earlier studies of $D^*$ decays. The agreement is, of course, welcome but not  unexpected and it supports the picture of compact tetraquarks bound by QCD forces. 
 
We find a strong dependence of decay rates from the pion momentum,  $\Gamma\propto q^5$. As a consequence, the decay  $Y(4230)\to \pi Z_c(4020)$ is strongly suppressed with respect to the decay into $\pi Z_c(3900)$, which does not seem to be supported by the cross sections reported by BES III. 
One possible explanation could be that the $Z_c(4020)$ events come from the second peak, $Y(4320)$.  A clarification of the distribution of $\pi D^* \bar D^*$ events in the region as  well as information on the decay modes of $Y(4320)$  would be very useful.

Production of exotic states in $e^+e^-$ annihilation goes essentially via $Y$ resonances. It is reasonable to assume that the  open strangeness state $Z_{cs}(3985)$ seen in~\cite{Ablikim:2020hsk}
\be
e^+e^-  \to K^+ Z^-_{cs}(3985)\to  K^+ (D_s^{*-}D^0+D_s^{-}D^{*0})\label{strangebes}
\ee
also arises from a $Y$-like resonance with $[cu][\bar c \bar u]$ or $[cs][\bar c \bar s]$ valence quark composition that decays to the final $K^+ Z_{cs}$ state by the elementary processes
\be
u\to  s ~K^+\quad {\rm or}\quad \bar s \to \bar u~ K^+
\ee 

If the hypothesis is correct, our analysis of K meson transitions,~ shows that strange members of the two nonets associated to $X(3872)$ and $Z_c(3900)$ should both appear in the final states of~\eqref{strangebes}, i.e. the $D_s\bar D^{*0} + c.c.$ spectrum should include as well the $Z_{cs}(4003)$ recently observed by LHCb in $B^+$ decay~\cite{Aaij:2021ivw} (the classification of the newly discovered $Z_{cs}$ resonances is considered in~\cite{Maiani:2021tri}). This is a crucial feature that can be tested in higher luminosity experiments.

\section{Production  and decay modes of $Y(4230)$ in $e^+e^-$ annihilation} \label{proddec}

A $J^{PC}=1^{--}$ resonance, $Y(4620)$,  was first observed by BaBar and confirmed by Belle  in  $e^+e^-$ annihilation with Initial State Radiation (ISR)\cite{Aubert:2005rm,Wang:2007ea}. BES III has later studied the $4620$ structure  with higher resolution  and shown that it is resolved in two lines, now indicated as $Y(4230)$ and $Y(4320)$ (see~\cite{changzheng}).

$Y$ states as $P$-wave tetraquarks have been described in \cite{Maiani:2014aja,Ali:2017wsf}. One expects four states $Y_1,\dots,Y_4$, the two lightest ones with spin composition 
\bea
&&Y_1= | (0, 0), L=1 \rangle_{J=1}\label{zerozero}\\
&&Y_2=\frac{1}{\sqrt{2}} \big( | (1, 0), L=1 \rangle + | (0, 1), L=1 \rangle \big)_{J=1}\label{onezero}
\eea
Valence quark composition is $[cq][\bar c\bar q]$, diquark and antidiquark spin are indicated  in parenthesis, $L$ is the orbital angular momentum.

It was noted in~\cite{Ali:2017wsf} that the mass difference of $Y_{1,2}$ arises from two contrasting contributions: the hyperfine interaction, which pushes $Y_1$ down, and the spin-orbit interaction, which pushes $Y_2$ down. We had chosen $M_1<M_2$ on the basis of a preliminary indication  that the $\gamma+X(3872)$ decay  was associated to $Y(4320)$, since this decay may arise from the $Y_2$ structure in Eq.~\eqref{onezero} and not  from  $Y_1$, Eq.~\eqref{zerozero}.

Later information~\cite{Ablikim:2019zio} indicates that the source of the $\gamma+X(3872)$ decay is instead $Y(4230)$. 
Consequently, we are led to change the assignment and propose $M_2<M_1$, that is
\be
 Y_2=Y(4230)\quad Y_1=Y(4320)~ {\rm or~higher}\label{present}
\ee

\begin{table}[htb!]
\centering
    \begin{tabular}{||c|c|c|c|c|c||} 
    \hline
 {\footnotesize {Ref.}} &{\footnotesize {$Z$(Mass)}} &{\footnotesize {$\sqrt{s}$~(GeV}) }& {\footnotesize {$e+ e\to Y(4230)\to\dots$}}& {\footnotesize{$Q$-value }}& {\footnotesize {$\sigma$~(pb) }}\\
 \hline
{\footnotesize { PRL 115\cite{Ablikim:2015gda}}}&{\footnotesize{$Z_c(3885)$}}&{\footnotesize{$4.226$}}&{\footnotesize {$\pi^0 Z_c^0\to \pi^0 (D  {\bar  D}^*+ c.c.)^0$}}&{\footnotesize{$197$}}&{\footnotesize{ $77\pm 21$}}\\
  \hline
 {\footnotesize {PRD 92\cite{Ablikim:2015swa}}}&{\footnotesize{$Z_c(3885)$}}&{\footnotesize{$4.23$}}&{\footnotesize {$[\pi^+ Z_c^- + c.c. ]\to[ \pi^+  (D\bar D^*)^- + c.c.]$}}&{\footnotesize{$197$}}&{\footnotesize {$141\pm 14$}}\\
 \hline
  {\footnotesize {PRL 112 \cite{Ablikim:2013emm}}}&{\footnotesize{$Z_c(4020)$}}&{\footnotesize{$4.26$}}&{\footnotesize {$[\pi^+ Z_c^- + c.c.]\to [\pi^+  (D^*\bar D^*)^- + c.c.]$}}&{\footnotesize{$65$}}&{\footnotesize{$\begin{array}{c}(0.65\pm 0.11)\cdot (137\pm 17)=\\ =89\pm19\end{array}$}}\\
 \hline
  {\footnotesize {PRL 115 \cite{Ablikim:2015vvn}}}&{\footnotesize{$Z_c(4020)$}}&{\footnotesize{$4.23$}}&{\footnotesize {$\pi^0 Z_c^0 \to \pi^0  (D^*\bar D^*)^0$}}&{\footnotesize{$65$}}&{\footnotesize {$62\pm 12$}}\\
  \hline
  {\footnotesize {PRL 115\cite{Ablikim:2013dyn}}}&{\footnotesize{$X(3872)$}}&{\footnotesize{$4.226$}}&{\footnotesize {$\gamma X \to \gamma  \pi^+\pi^-J/\psi$}}&{\footnotesize{$--$}}&{\footnotesize {$0.27\pm 0.09\pm0.12$}}\\
  \hline
  {\footnotesize {PRD100\cite{Li:2019kpj}}}&{\footnotesize{$$}}&{\footnotesize{$$}}&{\footnotesize {$ \gamma  X$}}&{\footnotesize{$354$}}&{\footnotesize {$5.5^{+2.8}_{-3.6}$}}\\
\hline {\footnotesize {2011.07855\cite{Ablikim:2020hsk}}}&{\footnotesize{$Z_{cs}(3982)$}}&{\footnotesize{$4.681$}}&{\footnotesize{$K^+Z_{cs}^- \to K^+ (D^{*-}_s D^{0}+ D^{-}_s D^{*0} )$}}&{\footnotesize {$199$}}&{\footnotesize{$4.4\pm 0.9$}}\\
\hline
\end{tabular}
 \caption{\footnotesize{$e^+e^-$ annhilation cross sections into  exotic hadrons, determined by BES III in the $Y(4230)$ region.}}
\label{data}
\end{table}
Table~\ref{data} summarizes the cross sections of different final states produced in $e^+ e^-$ annihilation at the $Y(4230)$ peak. Cross sections are related to the width $\Gamma(Y(4230)\to f)$ by the formula
\be
\sigma_{\rm peak}(e^+e^- \to Y(4230) \to f)=\frac{12\pi}{M_Y^2}\frac{\Gamma_e \Gamma_f}{\Gamma^2}
\ee
($\Gamma_e$ is the width to an electron pair). The total width of $Y(4230)$ is estimated in~\cite{changzheng}
\be
 \Gamma(Y(4230))=(56.0 \pm 3.6 \pm 6.9)~{\rm MeV}\label{totrate}
 \ee

Data from BES III indicate
that $Y(4230)$ is isoscalar~\cite{BESIII:2020pov, Collaboration:2017njt}. Thus, denoting by $Y_{u,d}$ the $Y_2$ states with $u\bar u$ and $d\bar d$ valence quarks, we  take
\be
Y(4230)=Y_2=\frac{Y_u+Y_d}{\sqrt{2}}\label{isoscalar}
\ee

\section{$Y(4230)$ transitions to $S$-wave tetraquarks}\label{ytrans}

We consider the decays
\bea
&& Y_2 \to \gamma+X \label{xgam}\\
&& Y_2 \to \pi+Z \label{zpi}\\
&& Y_2 \to \pi+Z^\prime \label{zprimpi}
\eea
where $X,~Z$ and $Z^\prime$ are the $S$-wave tetraquarks
\bea
X&=& \frac{1}{\sqrt{2}} \big( | (1, 0),~L=0 \rangle + | (0, 1),~L=0 \rangle \big)_{J=1}\label{splus} \\
Z&=& \frac{1}{\sqrt{2}} \big( | (1, 0),~L=0 \rangle - | (0, 1),~L=0 \rangle\big)_{J=1}\label{smin}\\
Z^\prime &=&   | (1, 1),~L=0 \rangle_{J=1}  \label{sminp}
\eea

The decay \eqref{xgam} as a dipole transition $L=1\to L=0,~ \Delta S=0$ has been considered in \cite{Chen:2015dig}. Here we rederive the result as an introduction to pionic transitions.

We work in the non-relativistic approximation and describe the states with wave functions in spin and coordinate space.
In the rest frame of $Y_2$
\be
|Y_2^a\rangle=N_Y\Big[\epsilon_{abc} S^{b(+)}\frac{\xi^c}{r} R_{2P}(r)\Big] \label{y2wf}
\ee
$N_Y$ is a normalization constant,  the spin wave function from~\eqref{onezero} and \eqref{splus} is
\be
S^{a(+)}=\frac{(c\sigma_2\sigma^a u)_{\bm x}(\bar c \sigma_2 \bar u)_{\bm y}+(c\sigma_2 u)_{\bm x}(\bar c \sigma_2\sigma^a \bar u)_{\bm y}}{2\sqrt{2}}\label{swfplus}
\ee
We indicate with a bar the charge-conjugate quark fields, ${\bm {x, y}}$ are diquark and antidiquark coordinates, ${\bm {\xi=x-y}}$ the relative coordinate and $r=|{\bm \xi}|$  the relative radius. The plus sign in $S^{a(+)}$ reminds of the charge conjugation, as defined on the rhs of~\eqref{swfplus}.

Considering the decay into $X$, we take
\be
|X^a \rangle=N_X ~\Big[S^{a(+)}~R_{1S}(r)\Big]
\ee
and normalize spin w.f. according to
\be
S^{a(+)}\cdot S^{b(+)}=\delta_{ab}
\ee 
Thus
\bea
\delta_{ab}&=&\langle Y_2^a|Y_2^b\rangle=N_Y^2~\delta_{ab} \frac{8\pi}{3} \int  dr ~y_{2P}(r)^2\notag\\
\delta_{ab}&=&\langle X^a|X^b\rangle=N_X^2~\delta_{ab}  4\pi \int dr~y_{1S}(r)^2\notag
\eea
with $y(r)=r R(r)$, and $R(r)$  the radial wave function.

{\bf \emph{Radiative decay.}}
~We work in the radiation gauge, $A^0=0$ and ${\bm \nabla}\cdot{\bm A}=0$. The photon couples to $u$ and to other quarks with the basic lagrangian \be
{\cal L}_{\rm e.m.}=e Q_u\, \bar u({\bm x}) \, {\bm {\gamma \cdot A}}({\bm x})\, u({\bm x}) + (u\to d)
\ee

The elementary transition amplitudes are 
\bea
{\cal M}_u&=&e Q_u\,\chi^\dag \left[\frac{({\bm p+\bm q})\cdot{\bm \sigma}}{2m_u}\sigma^a+\sigma^a\frac{{\bm {p}}\cdot{\bm \sigma}}{2m_u}\right]\chi\,\epsilon^a=e Q_u\,\chi^\dag\left[\frac{ {\bm {p}}_u{\bm{\cdot  \epsilon}}}{m_u} + \frac{i{\bm {q\wedge \epsilon \cdot \sigma}_u}}{2m_u}\right]\chi\label{transu}\\
{\cal M}_{\bar u}&=&-e Q_u\,\chi^\dag \left[\frac{({\bm p+\bm q})\cdot{\bm \sigma}}{2m_u}\sigma^a+\sigma^a\frac{{\bm {p}}\cdot{\bm \sigma}}{2m_u}\right]\chi\,\epsilon^a=- e Q_u\,\chi^\dag\left[\frac{ {\bm p}_{\bar u}{\bm {\cdot \epsilon}}}{m_{\bar u}} +\frac{i{\bm {q\wedge \epsilon \cdot \sigma}_{\bar u}}}{2m_u}\right] \chi\label{transubar}
\eea
The minus sign in ${\cal M}_{\bar u}$ arises from charge conjugation. In view of large mass denominators, we neglect radiation from the charm quarks. 

The right-hand sides of these equations contain products of operators acting on the spin and space wave functions of the initial tetraquark multiplied by variables of the electromagnetic field. As usual, we identify
\be
{\bm p}_u=-i{\bm \partial}_{\bm x}\qquad{\bm p}_{\bar u}=-i{\bm \partial}_{\bm y}
\ee
Acting on functions of ${\bm {\xi}=\bm {x-y}}$
\be
{\bm \partial}_{\bm x}={\bm \partial}_{\bm \xi}=i{\bm p}_{\bm \xi}\qquad {\bm \partial}_{\bm y}=-{\bm \partial}_{\bm \xi}=-i{\bm p}_{\bm \xi}
\ee
Further, we set 
\be
\frac{{\bm p}_{\bm \xi}}{m_u}={\bm {v}}=\frac{d{\bm \xi}}{dt}=i \omega {\bm \xi}\qquad \omega=|{\bm q}|
\ee
and the hamiltonian acting on tetraquark wave functions is
\bea
&&H_I=e Q_u~ \Big[{\bm {\xi\cdot E}} + \frac{1}{2m_u}\Big({\bm {\sigma}_u}-{\bm {\sigma}_{\bar u}}\Big)\cdot{\bm B}\Big]\label{hamgamma}
\eea
with ${\bm E}$ and ${\bm B}$ the electric and magnetic fields. The first term corresponds to the well known electric dipole transition that changes by one unit the orbital angular momentum, leaving the spin wave function unchanged~\cite{landau4}. One obtains
\bea 
 {\cal M}^{ab}(Y_2 \to \gamma + X)&=&e Q_u~\langle X^a| \xi^c|Y_2^b\rangle ~i\omega \epsilon^c= \epsilon_{abc}~ \epsilon^c~\frac{\omega}{\sqrt{6}}~\langle r \rangle_{2P\to 1S}\notag\\
 \langle r \rangle_{2P\to 1S}&=&\frac{\int dr~[ y_{1S}(r)~ r ~y_{2P} (r)]}{\sqrt{\int dr~ y_{1S}(r)^2 \int dr~y_{2P}(r)^2}}
\eea
with $\omega=M_Y-M_X=\omega_X $ and
\bea
&&\Gamma(Y_2 \to \gamma + X)= \frac{4\alpha}{9}~Q_u^2~\omega_X^3 \langle r \rangle_{2P\to 1S}^2
\eea

For isoscalar $Y(4230)$ we use~\eqref{isoscalar}. Summing incoherently over  the final states $X_u$ and $X_d$, see~\cite{Maiani:2017kyi}, we get
\be
Q_u^2 \to Q_{eff}^2=\frac{1}{2}(Q_u^2 + Q_d^2)=\frac{5}{18}\notag
\ee
and
 \bea
&&\Gamma(Y(4320 ) \to \gamma + X)=0.322~{\rm MeV}~(\omega_X~\langle r \rangle_{2P\to 1S})^2\label{gamzfin}
\eea
  
{\bf $\pi^0$ \emph{emission.}}
We assume that quarks couple to pions via the  isovector, axial vector current~\footnote{To our knowledge, the quark-pion, axial vector interaction to describe pionic hadron decays has been first introduced in~\cite{weisskopf}.}: 
\bea
&&{\cal L}^\pi_I=~\frac{g}{f_\pi}~\bar q \gamma^\mu \gamma_5(\partial_\mu{\bm \pi})q \label{lagpion}\\
&&{\bm \pi}=\frac{{\tau^i \pi^i}}{\sqrt{2}}=
\begin{pmatrix}\frac{\pi^0}{\sqrt{2}} & \pi^+\\\pi^- & - \frac{\pi^0}{\sqrt{2}}
\end{pmatrix}
\label{pionmatr}
\eea
We follow ~\cite{Casalbuoni:1996pg} for the definition of the coupling $g$ and 
\be 
f_\pi =132~{\rm MeV}
\ee
The lagrangian contains the time derivative of the pion field. Applying the Legendre transformation, the interaction  hamiltonian is
\be
{\cal H}^\pi_I=~\frac{g}{f_\pi}~ {\bm {A\cdot \nabla \pi}} \label{hampion}
\ee

The elementary quark transition is determined by 
\bea
&&{\bm {q\cdot }}{\bm A}_u=\bar u(p+q)\, {\bm {q \cdot\gamma }} \gamma_5 \, u( p)= [{\bm{q\cdot\sigma}}+\frac{q^2}{4m_u}\frac{\bm {p}}{m_u} \cdot {\bm  \sigma}+\dots]\label{axcurr}
\eea
The first term corresponds to $\Delta L=0$, operative in $D^* \to D \pi$~\cite{Casalbuoni:1996pg}, the second to $\Delta L=1$, for $Y$ and $D_1$ pionic decay, dots indicate terms with $\Delta L>1$.
Using charge conjugation symmetry, restricting to the $\Delta L=1$ term and specializing to the $\pi^0$ case, we obtain the hamitonian 
\be
H^\pi_I=\frac{g}{f_\pi}\frac{q^2}{4m_u}\, i\omega\, \chi^\dag[ {\bm {\xi \cdot}}({\bm \sigma}_{ u}-{\bm \sigma}_{\bar u}) - (u \to d)]\chi \, \frac{\pi^0}{\sqrt{2}} \label{hampi0}
\ee

We note the results of applying the spin operators to the components of the spin wave function
\bea
 { \sigma}^b_u(c\sigma_2 u)&=& (c\sigma_2 {\sigma}^bu)\notag\\
 { \sigma}^b_u(c\sigma_2  \sigma^a u)&=&(c\sigma_2\sigma^a{\sigma}^bu)=\Big[ \delta_{ab}(\bar c \sigma_2\bar u)+i\epsilon_{abc}(\bar c \sigma_2 \sigma^c \bar u)\Big]\notag \\
-\sigma_{\bar u}^b(\bar c\sigma_2 \bar u)&=&-(\bar c\sigma_2 \sigma^b{\bar u}) \notag\\
-\sigma_{\bar u}^b(\bar c\sigma_2  \sigma^a \bar u)&=&-(\bar c\sigma_2\sigma^a{\sigma}^b\bar u)=-\Big[\delta_{ab}(\bar c \sigma_2\bar u)+i\epsilon_{abc}(\bar c \sigma_2 \sigma^c \bar u)\Big]
\eea

Defining
\bea
S_u^{a(-)}&=&\frac{(c \sigma_2 \sigma^a  u)(\bar a \sigma_2\bar u)-(c\sigma_2 u)(\bar c \sigma_2 \sigma^a \bar u)}{2\sqrt{2}}=Z\label{essemen}\\
S_u^a&=&i\epsilon_{abc}\frac{(c\sigma_2\sigma^b u)(\bar c\sigma_2 \sigma^c \bar u)}{2\sqrt{2}}=Z^\prime \label{suno}
\eea
we obtain 

\bea
&&(\sigma_u^b-\sigma_{\bar u}^b) S_u^{a(+)}= i\epsilon_{abc}\Big[ S_u^{c(-)}+S_u^c \Big]\label{fin}
\eea

Note that,  going from $Y$ to $Z$ or $Z^\prime$, the minus sign between $\sigma_u$ and $\sigma_{\bar u}$ changes the charge conjugation sign of the spin w.f.. Similarly, the minus sign between the $u$ and $d$ term in \eqref{hampi0} changes the  $S_u$ and $S_d$ combination  from $I=0$ (in $Y$) to $I=1$  (in $Z$ and $Z^\prime$).

In conclusion, we find
\be
{\cal M}^{ab}=\langle Z^a|H^\pi_I|Y_2^b\rangle =\delta_{ab}\sqrt{\frac{2}{3}}\frac{g}{\sqrt{2}}~\left(\frac{q^2}{4f_\pi m_u}\right)~(\omega_Z \langle r\rangle_{2P\to 1S})\label{pionmatrel}
\ee
and 
\bea
\Gamma(Y(4230) \to Z^0\pi^0)&=&|{\cal M}^{11}|^2 \frac{1}{(2\pi)^3} ~4\pi\int \frac{q\omega d\omega}{2\omega}~(2\pi)\delta(\Delta M-\omega)=\frac{q}{2\pi}~|{\cal M}^{11}|^2 =\notag\\
&=&\frac{q}{6\pi}  g^2~\left(\frac{q^2}{4f_\pi m_u}\right)^2 \left(\frac{\omega_Z}{\omega_X}\right)^2~(\omega_X \langle r\rangle _{2P\to 1S})^2=4.36~{\rm MeV}~g^2  (\omega_X \langle r\rangle _{2P\to 1S})^2\label{gampionfin}
\eea
$q$ is the decay momentum, $\omega_Z=M_Y-M_Z$, we have chosen to normalize the radius with $\omega_X$, for comparison with Eq.~\eqref{gamzfin} and $m_u=308$~MeV, from the constituent quark model spectrum of mesons (see e.g.~\cite{Maiani:2004vq,Ali:2019roi}).

\section{Charge Conjugation in $Y$ and other tetraquark nonets}\label{chargeconjug}

A charge conjugation quantum number can be given to each self conjugate SU(3)$_f$ multiplet according to
\be
{\cal C}T{\cal C}=\eta_T {\tilde T}\label{cconj}
\ee
where ${\cal C}$ denotes the operator of charge conjugation, $T$ the matrix representing the multiplet in SU(3) space and ${\tilde T}$ the transpose matrix. $\eta_T$ is the sign taken by neutral members, but it can be attributed to all members of the multiplet. In the exact SU(3)$_f$ limit, $\eta$ is conserved in strong and electromagnetic decays. $\eta=-1$ is given to the electromagnetic current $J^\mu$ and to $Y^\mu$  while $\eta_{K,\pi}=+1$. 

We extend $Y_2$ to a full nonet that we write as (omitting the overall normalization for brevity)
\bea
Y_2(a,\bar b)^\gamma&=&  \epsilon_{\gamma\alpha \beta}~\Big[(D_a^\alpha \bar D_{\bar b})+(D_a \bar D_{\bar b}^\alpha)\Big]~ \frac{\xi^\beta }{r}~F(\xi)\notag\\
D_a^\alpha &=& (c\sigma_2\sigma^\alpha q_a); ~\bar D_{\bar b}=(\bar c\sigma_2 \bar q_b),~{\rm etc.}
\eea
$F(\xi)$ is the wave function in the relative coordinate, even under  $\xi \to -\xi$.

The lagrangian \eqref{lagpion} generalizes to
\be
{\cal L}^\pi_I=~\frac{g}{f_\pi}~\bar q \gamma^\mu \gamma_5( \partial_\mu{\bm M})q \label{lagmes}
\ee
where
\be
{\bm M}=\frac{{ \lambda}^i M^i}{\sqrt{2}}=
\begin{pmatrix}\frac{\pi^0}{\sqrt{2}}+\frac{\eta_8}{\sqrt{6}} & \pi^+ &{\footnotesize{K^+}}\\ \pi^- &-\frac{\pi^0}{\sqrt{2}}+\frac{\eta_8}{\sqrt{6}}& {\footnotesize{K^0}}\\{\footnotesize{K^- }}& {\footnotesize{\bar K^0}} &{\footnotesize{-2}} \frac{\eta_8}{\sqrt{6}}\end{pmatrix}
\ee
($\lambda^i$ are the Gell-Mann matrices). Correspondingly, the action on each diquark  of the hamiltonian  derived from \eqref{lagmes} is (for brevity, we omit two-dimensional spinors $\chi^\dagger$ and $\chi$,  which should bracket all the expressions below)
\bea
{\cal H}_I D_a^\alpha &\propto&  M_{a a^\prime}({\bm {\sigma\cdot \xi}})^{\alpha\beta} D_{a^\prime}^\beta= (M D_{\sigma \xi})_a^\alpha\notag\\
 {\cal H}_I \bar D_{\bar b}&\propto& -({\bm {\sigma\cdot \xi}})(M_{\bar b^\prime \bar b} \bar D)_{\bar b^\prime}=-({\bm {\sigma\cdot \xi}})\bar D_{\bar b^\prime}M_{\bar b^\prime \bar b}=-(\bar D_{\sigma \xi} M)_{\bar b},~{\rm etc.}\label{sigmaxi}\notag
\eea
and
\be
{\cal H}_I (D_a^\alpha \bar D_{\bar b})\propto (M D_{\sigma \xi})^{\alpha}_a\bar D_{\bar b}+D_a^\alpha (\bar D_{\sigma \xi} M)_{\bar b}
\ee
where
\be
(D_{\sigma \xi})^{\alpha}_a=(c\sigma_2 (\sigma\cdot \xi)q_a)\qquad (\bar D_{\sigma \xi})_{\bar b}=(\bar c\sigma_2 (\sigma\cdot \xi)\bar q_{\bar b})\notag
\ee
Explicitly, 
\bea
&&\frac{\xi^\beta}{r}~\Big[(D _{\sigma \xi})^{\alpha}_a \bar D_{\bar b}+(D _{\sigma \xi})_a \bar D^\alpha_{\bar b}\Big]=\frac{\xi^\beta}{r}\xi^\rho~\Big[(c\sigma_2\sigma^\alpha \sigma^\rho q_a)(\bar c \sigma_2 \bar q_{\bar b})+(c\sigma_2\sigma^\rho q_a)(\bar c\sigma_2\sigma^\alpha  q_{\bar b})\Big]=\notag\\
&&=\frac{\xi^\beta}{r}\xi^\rho~\Big[ \delta^{\alpha\rho}~(c\sigma_2 q_a)(\bar c \sigma_2 \bar q_{\bar b})+i\epsilon^{\alpha \rho\nu}(c\sigma_2 \sigma^\nu q_a)(\bar c \sigma_2 \bar q_{\bar b})+(c\sigma_2\sigma^\rho q_a)(\bar c\sigma_2\sigma^\alpha  q_{\bar b})\Big]\label{left}
\eea
Applying similar arguments to the second line of \eqref{sigmaxi}, we obtain
\bea
&&\frac{\xi^\beta}{r}~\Big[(D )^{\alpha}_a (\bar D _{\sigma \xi})_{\bar b}+(D)_a (\bar D_{\sigma \xi})^\alpha_{\bar b}\Big]=\frac{\xi^\beta}{r}\xi^\rho~\Big[(c\sigma_2\sigma^\alpha q_a)(\bar c \sigma_2\sigma^\rho  \bar q_{\bar b})+(c\sigma_2 q_a)(\bar c\sigma_2\sigma^\alpha\sigma^\rho q_{\bar b})\Big]=\notag\\
&&=\frac{\xi^\beta}{r}\xi^\rho~\Big[(c\sigma_2\sigma^\alpha q_a)(\bar c \sigma_2\sigma^\rho  \bar q_{\bar b})+\delta^{\alpha \rho} ~(c\sigma_2 q_a)(\bar c\sigma_2 q_{\bar b})+i\epsilon_{\alpha \rho \nu}(c\sigma_2 q_a)(\bar c\sigma_2\sigma^\nu q_{\bar b})\Big]\label{right}
\eea
The expressions in \eqref{left} and \eqref{right} are to be integrated with functions symmetric under $\xi^a\to-\xi^a$, so we can replace $\xi^\beta \xi^\rho \to  \delta^{\beta\rho}~ r^2/3$. In addition, in the square brackets we can add and subtract terms that reconstruct the spin wave functions of  tetraquarks of charge conjugation $+1$, spin $0, 1, 0, 2$,
namely $X, ~X_0, ~X^\prime_0, X_2$, and of charge conjugation $-1$, spin $1$, i.e. $Z,~Z^\prime$. Indicating for brevity only $X,~Z,~Z^\prime$, Eqs.~\eqref{splus}~\eqref{smin} and \eqref{sminp}, we obtain
\bea
&&{\cal H}_I~Y_2\propto  \frac{r}{3}~i\epsilon_{\alpha\beta\nu}~\frac{1}{2}\Big[(MX^\nu-X^\nu M)+(MZ^\nu+Z^{\nu}M) +(MZ^{\prime\nu}+ Z^{\prime\nu}M)+\dots\Big]
\eea
Multiplying by the $SU(3)$ matrix representing $Y$  and taking the trace we obtain the exact SU(3)$_f$ rules for the couplings of a $C=-1$ vector nonet to $M$ plus an $S$-wave tetraquark of charge conjugation $\eta_T$: 
\bea
{\cal H}_I &\propto& {\rm Tr}[Y[M,X]]
\quad (\eta_X=+1)\notag\\
{\cal H}_I &\propto& {\rm Tr}[Y\{M,Z\}] \quad (\eta_Z=-1).
\eea
In particular, for $Y(4230)$: $Y={\rm diag}(1/\sqrt{2},1/\sqrt{2},0)$, we obtain vanishing coupling $Y\to \pi X$ and Eq.~\eqref{pionmatrel} for $Y_2\to \pi^0Z^0$.

Summarizing, we obtain the selection rules:
\begin{enumerate}
\item  $Y_2 (I=0)$ does not decay into $\pi^{\pm,0} X^{\mp,0}$ 
\item $Y_2 (I=1)$ decays into $\pi^+ X^- - \pi^- X^+$
\item $Y_2(I=0)$   decays into $\pi^+Z^-+\pi^- Z^++\pi^0Z^0$, same for $Z^\prime$
\item $Y_2(I=0,{\rm or}~1)$ or $Y[cs\bar c\bar s]$ all decay into $(K^+ X_{cs}^- ~-~ c.c.)$ and $(K^+ Z_{cs}^- ~+ ~c.c.)$
\item The decay $Z_{cs}\to J/\psi \, K$ is allowed in the exact SU(3)$_f$ limit with
\be
{\cal H}_I=\lambda \mu ~\psi ~({\rm Tr}\{Z,M\}),~[\mu]={\rm mass}
\ee
\item The decay $X_{cs}\to J/\psi\,  K$ may occur to first order in SU(3)$_f$ symmetry breaking with
\be
{\cal H}_I=\lambda~i\psi ~{\rm Tr}([\epsilon_8[X,M])\sim \lambda~ (m_s-m_u)~i \psi (X_{cs}^+K^-~-~c.c.)
\ee
\end{enumerate}

\section{Radiative and pionic decays: ${{ D^*}}$ and ${ {D_1}}$ mesons}
\label{ddecays}

{\bf\emph{ The decay}}  $\bm{D^{*}\to \gamma+D}$. In its spin dependent part, the hamiltonian~\eqref{hamgamma}  describes the radiative decay of $D^*$,  $\Delta S=1$ and no change in orbital angular momentum. Setting the charm quark in the origin, $D^*$ and $D$ are represented by
\bea
&&D^{(*a)}={\bm V}^{(a)}\cdot \Big(c^\dagger(0) \frac{{\bm \sigma}}{\sqrt{2}}u({\bm x})\Big)~R(|{\bm x}|)\qquad D=\Big(c^\dagger(0) \frac{1}{\sqrt{2}} u({\bm x})\Big)~R(|{\bm x}|)\notag\\
&&{\cal M}(D^{*0}\to \gamma+ D^0)=\frac{eQ_u}{2m_u} \chi^\dag\, {\bm V}^{(a)}\cdot {\bm {q\wedge \epsilon}^{(b)}}\, \chi\notag
\eea
and we obtain
\bea
&& \Gamma(D^{*0}\to \gamma+ D^0)=\frac{\alpha}{3}~ \left(\frac{Q_u}{m_u}\right)^2~  q^3\label{dstardec}
\eea
Ref.~\cite{Casalbuoni:1996pg} can be consulted for a discussion of the $D^*$ decay rate and  the strong interaction corrections to \eqref{dstardec}.

{\bf\emph{ The decay}} ${\bm {D^{*+} \to \pi^0 D^+}}$.
From the hamiltonian \eqref{hampion} and Eq.~\eqref{axcurr}, the relevant term in the hamiltonian is
\be
{\cal H}_I^\pi=\frac{g}{\sqrt{2}f_\pi}\, {\bm {q \cdot \sigma_u}}\notag
\ee
so that
\be
{\cal M}(D^{*+} \to \pi^0 D^+)=\frac{g}{\sqrt{2}f_\pi}~ {\bm {V \cdot q}} \notag
\ee
and
\bea
&& \Gamma(D^{*+} \to \pi^0 D^+)=
g^2~\frac{(p^0)^3}{12\pi f_\pi^2}\notag \label{gamdstpi0}
\eea
with $p^0$ the decay momentum. Also
\be
  \Gamma(D^{*+} \to \pi^+ D^0)=g^2\frac{(p^+)^3}{6\pi f_\pi^2}\notag
  \ee
(decay momentum $p^+$). We reproduce the results of~\cite{Casalbuoni:1996pg}. We assume $D^{*+}$ decay to be dominated by $\pi D$ final states and use the $D^{*+}$ total width~\cite{pdg} to estimate the value of $g$, obtaining
\be
g\sim 0.56.\label{gdstar}
\ee
${\bm {D^0_1 \to \pi^0~D^{*0}}}$\, {\bf \emph{transition}}. 
$D_1(2420)$ is a well identified $P$-wave, positive parity charmed meson with total spin and angular momentum $S=J=1$. We can use its decay into $D^* \pi$ to calibrate the $\Delta L=1$ hamiltonian~\eqref{hampi0}. In analogy with~\eqref{y2wf}, we write the $D_1$ wave function as:

\bea
|D_1^a\rangle&=&N_1\Big[\epsilon_{abc} \frac{\xi^b}{r}~(\bar c\sigma_2\frac{\sigma^c}{\sqrt{2} }u)~R_{2P,D}(r)\Big] \label{d1wf}\\
\delta^{ab}&=&\langle D_1^a|D_1^b\rangle= N_1^2~\frac{8\pi}{3}~\int~dr~y_{2P,D}(r)^2\notag
\eea
where the subscript $D$ indicates that the QCD couplings of the $\bar c u$ system are used.

The decay is induced by the ${\bm \xi}$ dependent part of the hamiltonian, restricted to the $u$ term. Proceeding as before, we find
\be
{\cal M}^{ab}=\langle D^{*a}|H^\pi_I |D_1^b\rangle=\delta_{ab}~\frac{g}{\sqrt{2}}~\left(\frac{q^2}{4 f_\pi m_u}\right)\sqrt{\frac{2}{3}}\, (\omega \langle r\rangle_{2P,D \to 1S,D})\notag
\ee
and
\bea
\Gamma(D^0_1 \to \pi^0+D^{*0})&=&\frac{q_1}{6\pi}~ g^2 \left(\frac{q^2}{4 f_\pi m_u}\right)^2\, (\omega_1 \langle r\rangle_{2P,D\to 1S,D})^2\label{gamd1}\\
 \Gamma(D^0_1 \to \pi+D^{*})&=&3~\Gamma(D^0_1 \to \pi^0+D^{*0})
\eea
$q_1$ is the decay momentum, $\omega_1=M_{D_1}-M_{D^*}$ and we have assumed that the $\pi D^*$ modes saturate the total width. The transition radius in Eq.~\eqref{gamd1} is computed in the next Section, see Tab.~\ref{radius}. Using the experimental width~\cite{pdg} we find:
\be
g= 0.63\pm 0.08\label{gd1}
\ee
the error is estimated from the $D_1^0$ and $D_1^\pm$ width errors and variations in the estimated radius.

\section{Transition radius}\label{radius}

The transition radius for a diquarkonium was estimated  in~\cite{Chen:2015dig}, from the radial wave functions of a diquark-antidiquark system in a confining, QCD potential.  
 We solve numerically the two body, radial Schr\"odinger equation~\cite{code} with potential and diquark mass
\bea
&& V(r)=-\frac{\alpha_s}{r} + k r\qquad M_{[cq]}= 1.97~{\rm GeV}
\eea
Couplings are taken from  lattice calculation of charmonium spectrum~\cite{lattice}
\be
\alpha_s=0.3\qquad k=0.15~{\rm GeV}^2 \label{ourvalues}
\ee
Alternatively, Ref.~\cite{Chen:2015dig} uses the parameters of the Cornell potential~\cite{cornell}  or a pure confinement case:
\bea
&& \alpha_s=0.47\qquad k=0.19~{\rm GeV}^2~({\rm Cornell}) \label{corpot} \\
&&\alpha_s=0\qquad k=~0.25~{\rm GeV}^2~({\rm confinement~only} )\label{confpot}
\eea

For the $D_1\to D^*$ transition, we use the same potentials and $M_c=1.7,~m_u=0.308$~GeV.

\begin{table}[htb!]
\centering
    \begin{tabular}{||c|c|c|c||} 
    \hline
{\footnotesize{ Transition  $2P\to 1S$}}&{\footnotesize{lattice, Eq.~\eqref{ourvalues}}}& {\footnotesize{Cornell,~Eq.~\eqref{corpot}}} & {\footnotesize{pure confinement, Eq.~\eqref{confpot}}}\\
\hline
$Y(4230)\to X(3872)/Z_c(3900)$ &  $2.17$  &  $1.84$  & $2.15$\\
\hline
$D_1(2420)\to D^*$ &  $3.74$  &  $3.34$  & $3.36$\\
\hline
\end{tabular} 
 \caption{\footnotesize{Values of the transition radius $2P\to 1S$,  GeV$^{-1}$, for $P$-wave   tetraquark  and $D_1$.}}
\label{radius}
\end{table}
Results are reported in Tab.~\ref{radius}

\newpage

\section{Summary}\label{discuss}

{\bf\emph{The value of g.}}~The ratio of \eqref{gampionfin} to \eqref{gamzfin} depends on $g^2$ only:
\be
 R_\Gamma=\frac{\Gamma(Y4230)\to \pi^0 Z_c(3900))}{\Gamma(Y(4230)\to \gamma X(3872))}=13.5~g^2 
 \ee
  Assuming that $\Gamma(Z_c(3900)^0\to (D^* \bar D+ c.c)^0)$ saturates the $Z_c(3900)$ width, we obtain $g$ by comparison  to the ratio of the corresponding cross sections, Tab.~\ref{data}: 
 \be
R_\sigma=\frac{\sigma(e^+e^-\to  \pi^0Z_c(3900)^0 \to \pi^0 (D  {\bar  D}^*+ c.c.)^0)}{\sigma(e^+e^-\to \gamma X(3872))}=14 
\ee
at $\sqrt{s}=4.226$~GeV. From $R_\Gamma=R_\sigma$, taking into account the errors of the cross sections, we find
\be
g=1.0^{+0.6}_{-0.3}\label{gtetra}
\ee
that compares well with with $g=0.6-0.7$ obtained in~\eqref{gdstar} and \eqref{gd1}.

{\bf\emph{The $Z_c(4020)$ puzzle.}} The axial $\Delta L=1$ transition amplitude has a strong dependence from the pion momentum which reflects in a steep dependence of the rate: $\Gamma\propto q^5$ , see Eq.~\eqref{gampionfin}. The pion momentum of $Y(4230)\to \pi Z_c(4020)$ implies a suppression factor $\sim 30$ with respect to $Y(4230)\to \pi Z_c(3900)$, which does not seem to be supported by the cross sections in Tab.~\ref{data}. 

Would it be possible that the $Z_c(4020)$ events come from the second peak of the structure, $Y(4320)$? A clarification of the source of $D^* \bar D^*$ events in the region and of the decay modes of $Y(4320)$  would be very useful.

{\bf\emph{The total $Z_c(3900)$ width.}} Following \eqref{gtetra} and the transition radius in Tab.~\ref{radius} (lattice value), we estimate  the total rate 
\be
\Gamma(Y(4230)\to \pi Z_c(3900))=8^{+10}_{-4}~{\rm MeV}
\ee
corresponding to a fraction $(5-36)\%$ of the total $Y(4230)$ rate, Eq.~\eqref{totrate}.

\section*{Acknowledgements}

We acknowledge a very informative exchange with Chang-Zheng Yuan on the BES III data reported in Table \ref{data}.

\end{document}